\documentclass{webofc}
\usepackage[varg]{txfonts}   % Web of Conferences font

\usepackage{epstopdf}
\def\bea{\begin{eqnarray}}
\def\eea{\end{eqnarray}}

\def\pp{\mbox{$p$-$p$}}

\def\pa{\mbox{$p$-A}}

\def\auau{\mbox{Au-Au}}

\def\pbpb{\mbox{Pb-Pb}}
\def\ppb{\mbox{$p$-Pb}}
\def\pn{\mbox{$p$-N}}
\def\aa{\mbox{A-A}}

\def\pt{$p_t$}

\def\yt{$y_t$}

\def\nch{$n_{ch}$}
\def\mmpt{$\bar p_t$}

\begin{document}
\title{QGP droplet formation in small 
	asymmetric collision systems}

\author{\firstname{Thomas A.} \lastname{Trainor}\inst{1}
	\thanks{\email{ttrainor99@gmail.com}}\
}

\institute{University of Washington, Seattle, USA
          }

\abstract{%
The journal Nature recently published a letter titled "Creating small circular, elliptical, and triangular droplets of quark-gluon plasma" [1]. The basis for that claim is a combination of measured Fourier amplitudes $v_2$ and $v_3$ from collision systems $p$-Au, $d$-Au and $h$-Au (helion $h$ is the nucleus of atom $^3$He), Glauber Monte Carlo estimates of initial-state transverse collision geometries for those systems and hydro Monte Carlo descriptions of the $v_n$ data. Correspondence between hydro $v_n$ trends and data trends is interpreted as confirmation of ``collectivity'' occurring in the small collision systems, further interpreted to indicate QGP formation. QGP formation in small systems runs counter to pre-RHIC theoretical assumptions that QGP formation should require large collision systems (e.g.\ central A-A collisions). There is currently available a broad context of experimental data from \pp, \aa\ and \ppb\ collisions at the RHIC and LHC against which the validity of the Nature letter claims may be evaluated. This talk provides a summary of such results and their implications.

\noindent
[1] Nature Phys. 15, no. 3, 214 (2019).}
\maketitle

\section{Introduction}
\label{intro}

A recent letter published in the journal Nature reports observation of ``short-lived QGP droplets'' in 200 GeV $p$-Au, $d$-Au and $h$-Au ($h$ representing the helion nucleus of $^3$He)~\cite{nature}. The claimed observation is based on a combination of Glauber Monte Carlo estimates of initial conditions (IC), hydro theory evolution from the IC to the hadronic final state and comparison of theory results with measurements of azimuth Fourier components $v_2(p_t)$ and $v_3(p_t)$. The letter concludes: ``...hydrodynamical models which include QGP formation provide a simultaneous and quantitative description of the data in three systems.'' The overall argument is based on five critical assumptions that have been challenged in Ref.~\cite{tomnature}. In  this presentation I confront the five assumptions with evidence from a broad array of published data that contradict them. The context for comparison is the two-component (soft + hard) model (TCM) of hadron production in high-energy nuclear collisions. A third component inferred from measurements of a nonjet azimuth quadrupole structure is also introduced.

%%%%%%%%%%%
\section{Is a perfect liquid formed in A-A collisions?}
\label{sec-2}

Arguments in favor are based on azimuthal asymmetries ($v_2$, $v_3$ etc.) and jet quenching ($R_{AA}$ high-\pt\ suppression) as common manifestations of a dense and flowing QCD medium or QGP in \aa\ collisions. Reference~\cite{nature} argues by analogy that if some or all of those phenomena are observed in small asymmetric collision systems then QGP may be formed there as well. 

\begin{figure}[h]
	\includegraphics[width=1.25in,height=1.2in]{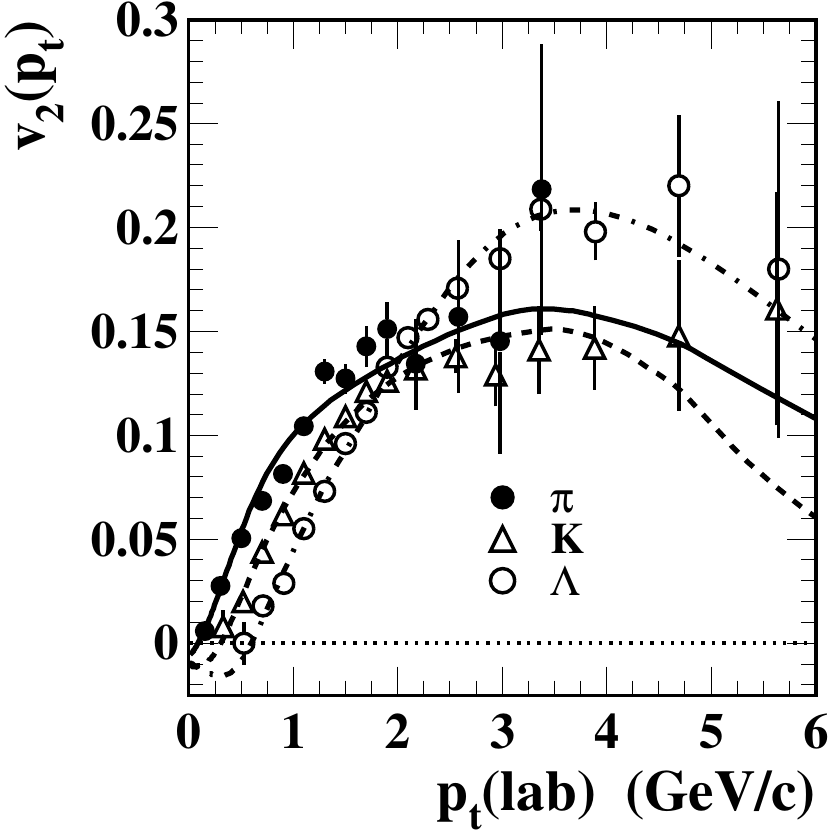}
	\includegraphics[width=1.25in,height=1.2in]{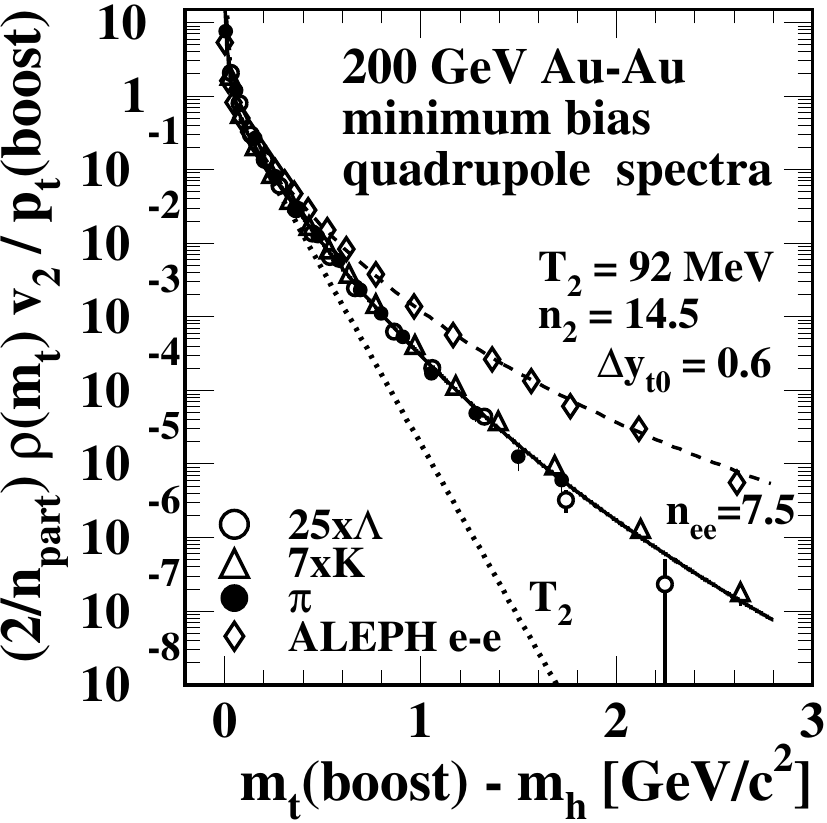}
	\includegraphics[width=1.25in,height=1.2in]{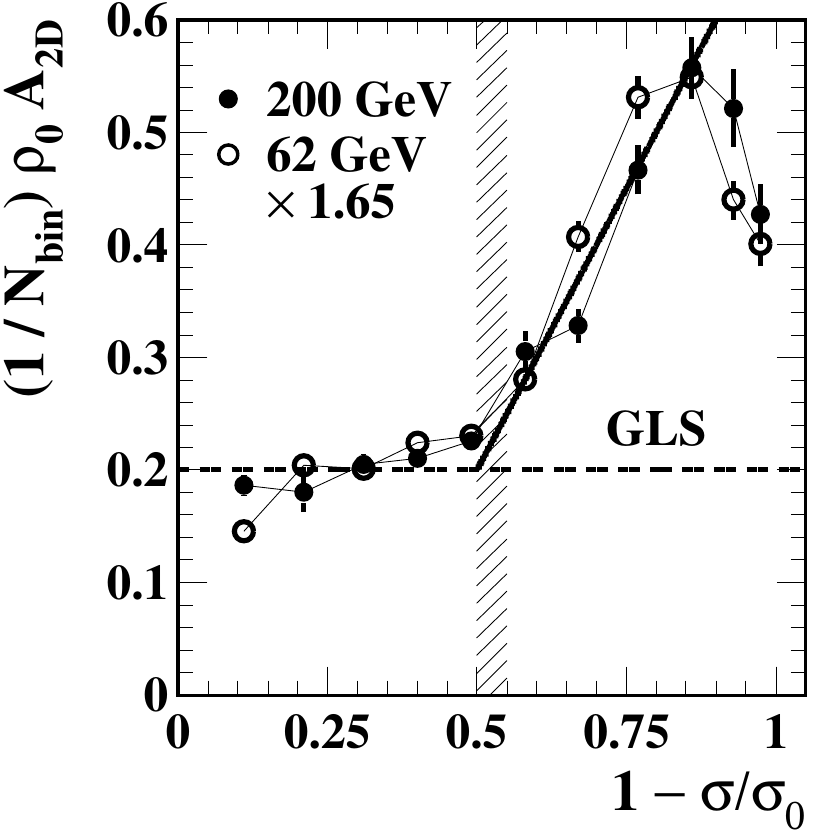}
	\includegraphics[width=1.25in,height=1.2in]{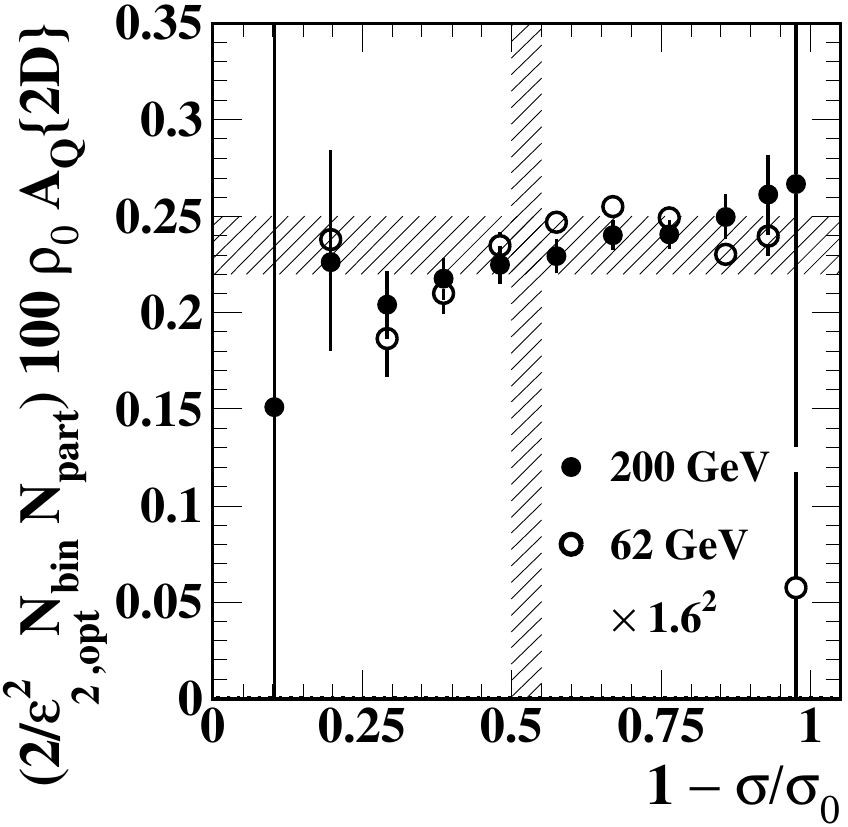}
	\caption{
		Left:  PID $v_2(p_t)$ data on $p_t(\text{lab})$ and corresponding quadrupole spectra on $m_t(\text{boost})$   from 200 GeV \auau\ collisions~\cite{quadspec}.
		Right: Centrality trends for jet quenching and $v_2$ from 200 GeV \auau~\cite{nonjetquad}.
	}
	\label{fig-1}     
\end{figure}

Figure~\ref{fig-1} (left) shows identified-hadron (PID) $v_2(p_t)$ for pions, kaons and Lambdas from 200 GeV \auau\ collisions. The conventional wisdom is that {\em mass ordering} at lower \pt\ indicates the presence of radial flow. However, a simple sequence of transformations takes those data from the first  to the second panel~\cite{quadspec}. Resulting {\em quadrupole spectra} for three species plotted in the boost frame fall on the same L\'evy distribution with slope parameter $T = 92$ MeV. The transformations assume a fixed boost $\Delta y_{t0} = 0.6$ from lab to boost frame. That result falsifies the notion that most hadrons experience Hubble expansion of a bulk medium. The small minority of hadrons ``carrying'' the nonjet quadrupole have a very different spectrum structure and experience a fixed boost corresponding to an expanding thin shell.

%%%%%%%%%%
%%%%%%%%%%

Figure~\ref{fig-1} (right) shows a comparison (based on 200 GeV \auau\ 2D angular correlations) between jet modification (third) exhibiting a {\em sharp transition} near 50\% centrality (hatched band) and the nonjet quadrupole trend (fourth) showing no deviation from a fixed trend from peripheral to central collisions~\cite{nonjetquad}. If there were a common dense QCD medium or QGP, with onset near  the sharp transition as suggested by jet modification, it is reasonable to expect that elliptic flow as manifested by $v_2$ data should show a corresponding marked change in its centrality trend, but no correspondence is observed. In other analysis the jet contribution to \pt\ spectra is described accurately by pQCD for all \auau\ centrality, albeit fragmentation functions are modified in more-central collisions~\cite{ppquad,hardspec}. That combination, consistent with other evidence, casts strong doubts on formation of a ``perfect liquid'' in \aa\ collisions.

%%%%%%%%%%%
\section{Is a Monte Carlo Glauber valid for $\bf x$-A collision geometry?}
\label{sec-3}

The hydro model utilized in Ref.~\cite{nature} requires estimation of the IC (i.e. the initial transverse geometry) for small asymmetric $x$-A collision systems. It is assumed that the Glauber Monte Carlo (MC) provides meaningful estimates of the IC for such systems. That assumption can be strongly questioned based on analysis of ensemble \mmpt\ data from 5 TeV \ppb\ collisions.

\begin{figure}[h]
	\includegraphics[width=1.25in,height=1.18in]{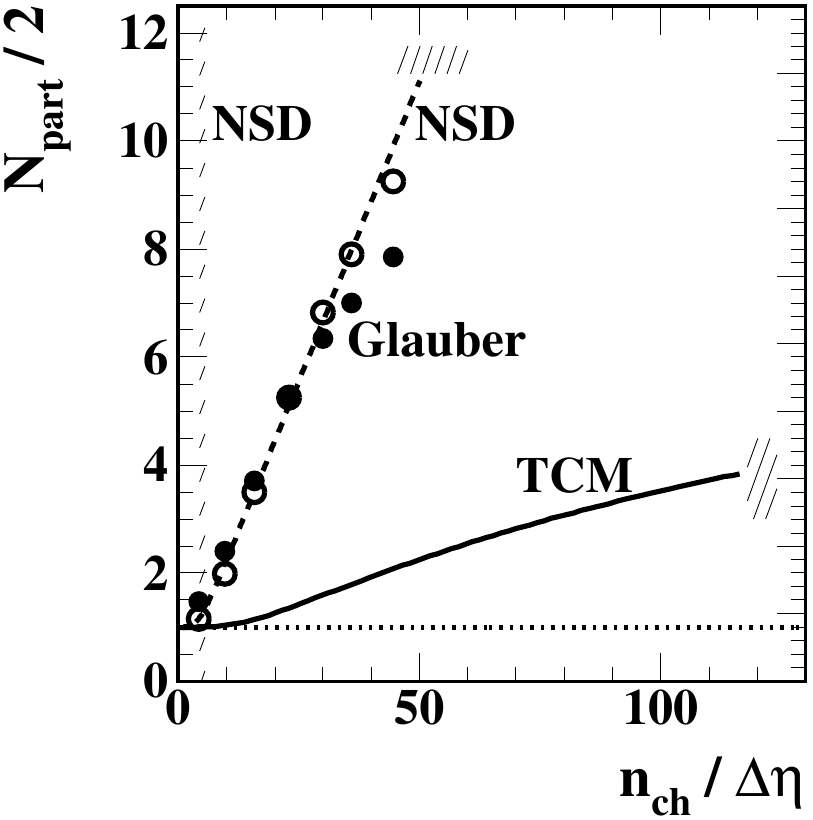}
	\includegraphics[width=1.25in,height=1.2in]{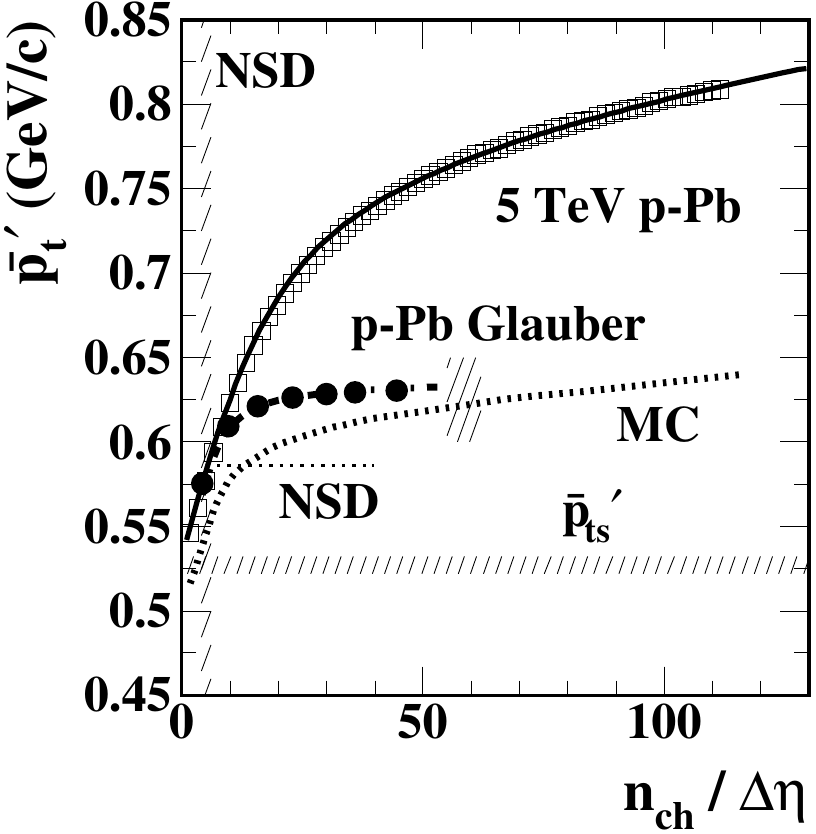}
	\includegraphics[width=1.25in,height=1.22in]{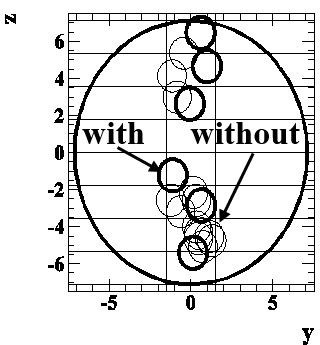}
	\includegraphics[width=1.25in,height=1.18in]{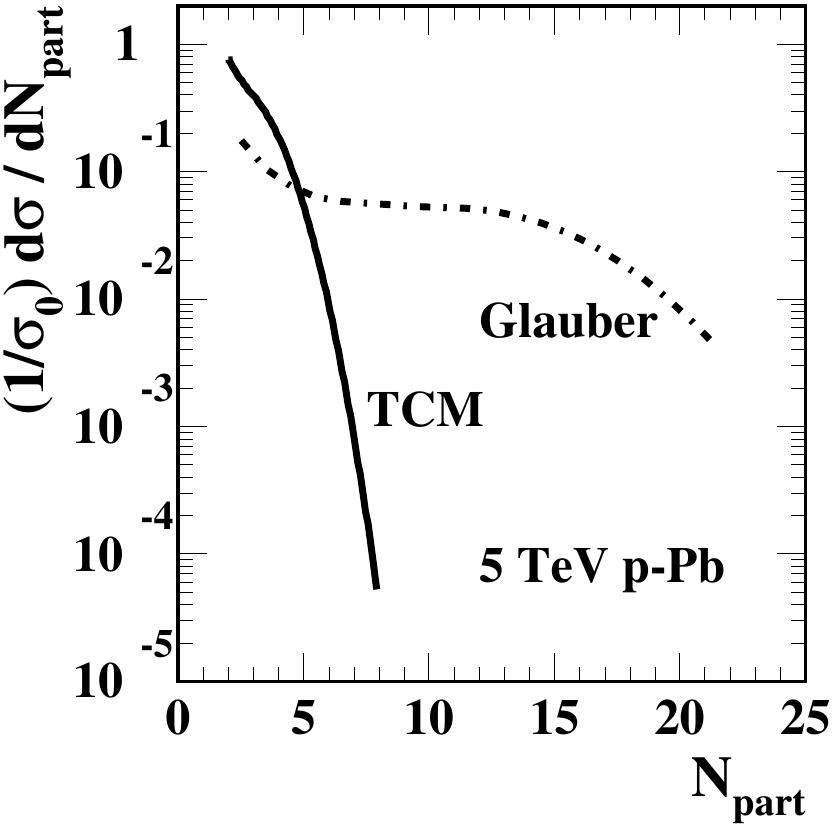}
	\caption{TCM vs Glauber geometry for 5 TeV \ppb\ collisions. 
		Left: Estimates of $N_{part}/2$ and \mmpt\ vs mean charge density $\bar \rho_0 = n_{ch} / \Delta \eta$~\cite{ppbpid,tomglauber}.
		Right:  Monte Carlo Glauber simulation of \ppb\ collisions~\cite{tomexclude}.
	}
	\label{fig-2}     
\end{figure}

Figure~\ref{fig-2} (left) compares MC Glauber estimates to TCM estimates for nucleon participant-pair number $N_{part}/2$ (first) and ensemble \mmpt\ (second) vs mean charge density $\bar \rho_0 = n_{ch} / \Delta \eta$ from 5 TeV \ppb\ collisions~\cite{ppbpid,tomglauber}. The MC Glauber estimates are based on the assumption that \pn\ collisions within \ppb\ collisions are on average equivalent to isolated non-single-diffractive (NSD) \pp\ collisions. In contrast, TCM estimates are based on an analysis of \mmpt\ data which indicates that \pn\ mean multiplicities vary strongly with \ppb\ centrality. The MC Glauber estimates (solid points) dramatically fail to describe \mmpt\ data (open squares) in the second panel whereas the TCM description (solid curve) describes the data accurately.
  
%%%%%%%%%%
%%%%%%%%%%

Figure~\ref{fig-2} (right) provides an explanation for the discrepancy. The third panel shows a Glauber MC simulation of a central \ppb\ collision. The total number of small circles denotes the number of geometric {\em encounters} of the projectile proton with target nucleons. The bold circles denote  actual \pn\ {\em collisions} allowed by an ``exclusivity'' condition: the projectile proton can interact with only {\em one nucleon at a time}~\cite{tomexclude}. The MC Glauber {\em with} exclusivity (labeled TCM in last panel) is consistent with \mmpt\ data and other aspects of \ppb\ data. The MC Glauber {\em without} exclusivity  produces severely biased IC estimates for $x$-A collisions.

%%%%%%%%%%%
\section{Do A-B $\bf p_t$ spectra provide evidence for radial flow?}
\label{sec-4}

Radial flow is conventionally inferred from \pt\ spectra by qualitative observations of increasing spectrum ``hardening'' (increased slope parameter) with increasing collision centrality and hadron mass, and by quantitative application to spectra of a blast-wave fit model with parameters $T_{kin}$ and $\bar \beta_t$, the latter interpreted as a quantitative measure of radial flow.

\begin{figure}[h]
	\includegraphics[width=1.25in,height=1.25in]{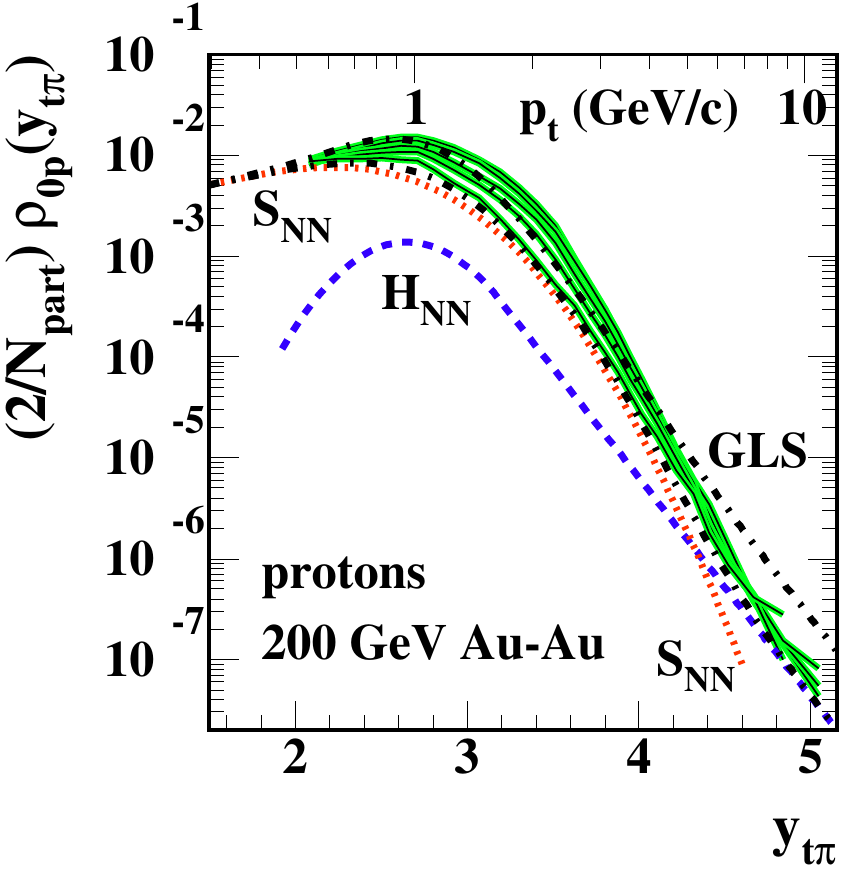}
	\includegraphics[width=1.25in,height=1.18in]{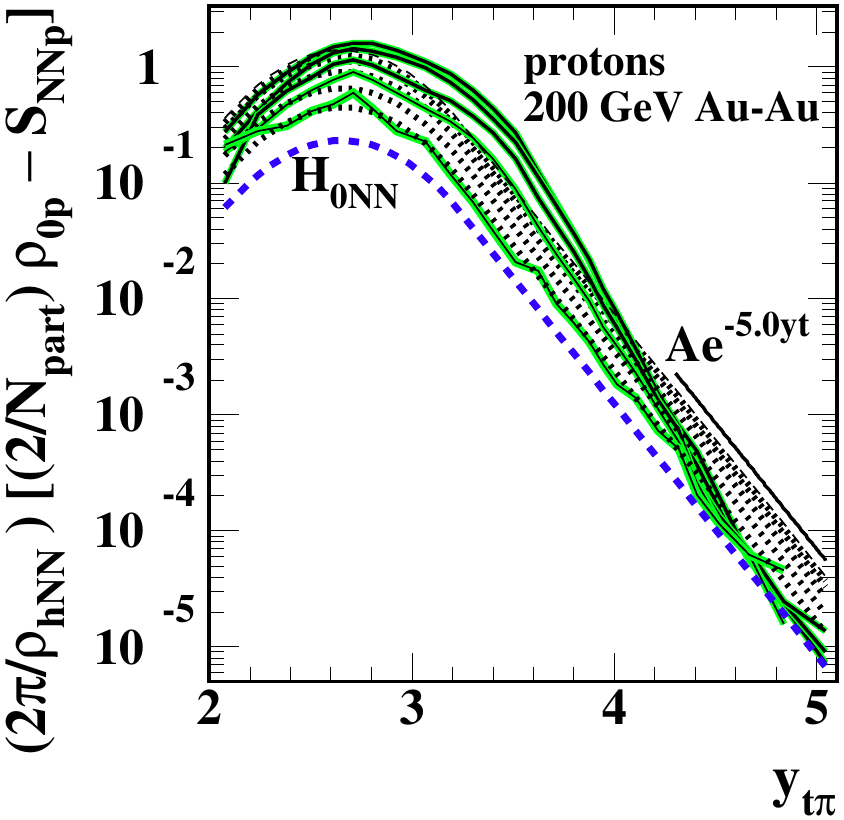}
	\includegraphics[width=2.5in,height=1.2in]{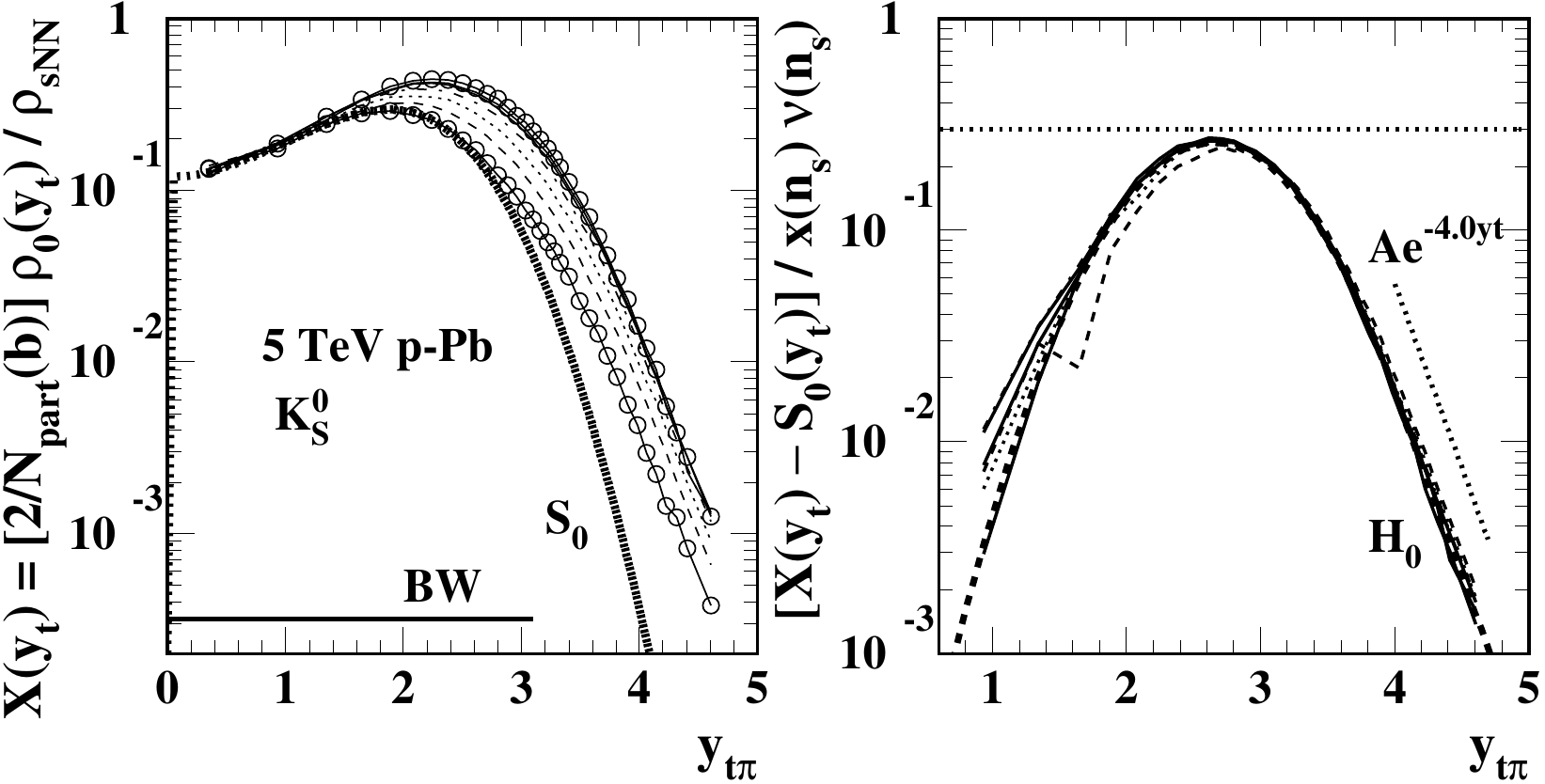}
	\caption{
		Left: Proton \yt\ spectrum TCM for 200 GeV \auau\ collisions~\cite{hardspec}.
		Right: $K^0_\text{S}$ \yt\ spectrum TCM for 5 TeV \ppb\ collisions~\cite{ppbpid}.
	}
	\label{fig-3}     
\end{figure}

Figure~\ref{fig-3} (left) shows a TCM analysis of proton spectra from 200 GeV \auau\ collisions~\cite{hardspec}. The first panel shows the full spectra while the second panel shows the extracted spectrum hard components (complete minimum-bias jet contribution). The spectra show no evidence for radial flow (boost or translation of the spectrum soft component on \yt\ leading to {\em suppression} at lower \yt). The hard components (second panel) indicate suppression at higher \yt\ (consistent with $R_{AA}$ and ``jet quenching'') but substantial {\em enhancement} near $y_t \approx 3.5$ compared to the TCM references (dotted). The hard components below $y_t = 3$ show no change with centrality. These \auau\ proton spectra thus provide no evidence for radial flow.

%%%%%%%%%%
%%%%%%%%%%

Figure~\ref{fig-3} (right) shows $K_\text{s}^0$ spectra from 5 TeV \ppb\ collisions extending down to zero momentum (full spectra and jet-related hard components respectively)~\cite{ppbpid}. If significant radial flow played a role there should be substantial suppression at lower \pt\ as a consequence of spectra boosted (translated to the right) on \yt. These spectra show no evidence for radial flow. The spectra above 0.5 GeV/c (\yt\ = 2) are dominated  by the minimum-bias jet contribution, and the jet contribution (fourth) shows no evidence for jet modification (changes in shape).

%%%%%%%%%%%
\section{Do Fourier amplitudes $\bf v_n$ measure flows?}
\label{sec-5}

Fourier amplitudes $v_n$ inferred from two-particle angular correlations can be derived from direct model fits to full 2D angular correlations or from Fourier fits to 1D azimuth projections of 2D angular correlations. In the latter case there is the possibility that certain jet-related structures when projected will contribute substantially to inferred $v_n$ as a ``nonflow'' bias.

\begin{figure}[h]
	\includegraphics[width=1.25in,height=1.2in]{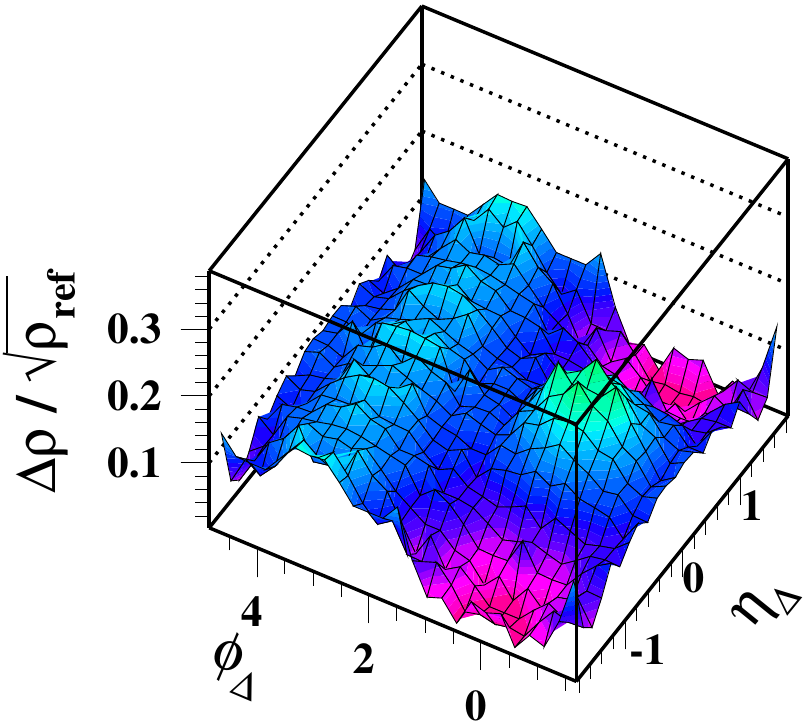}
	\includegraphics[width=1.25in,height=1.2in]{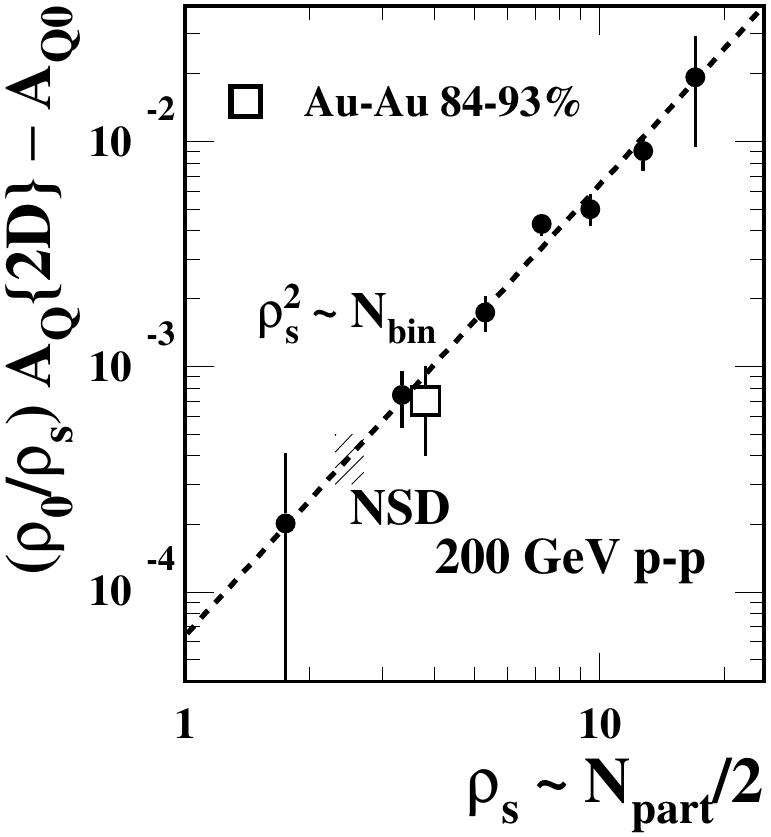}
	\includegraphics[width=1.25in,height=1.2in]{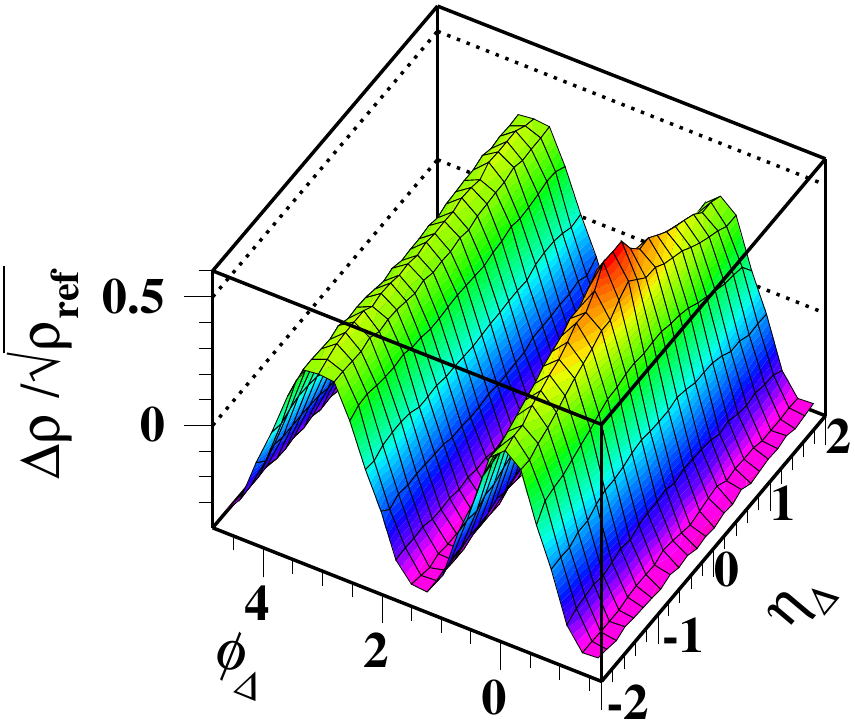}
	\includegraphics[width=1.25in,height=1.25in]{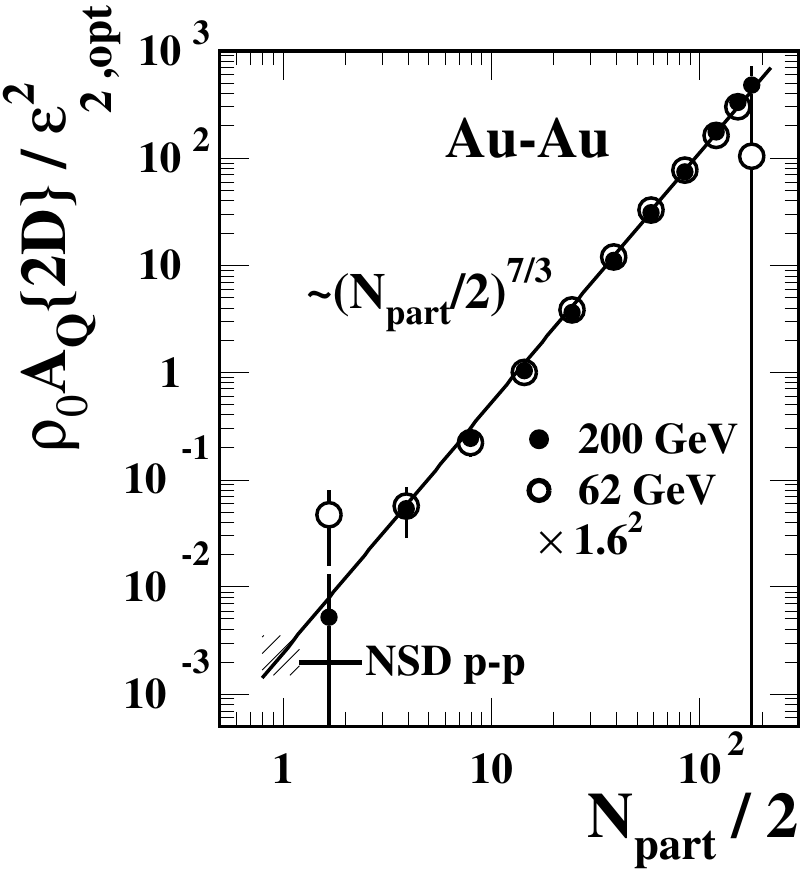}
	\caption{
		Left: Nonjet quadrupole \nch\ trend for 200 GeV \pp\ collisions~\cite{ppquad}.
		Right: Nonjet quadrupole centrality trend for 200 GeV \auau\ collisions~\cite{anomalous,v2ptb}.
	}
	\label{fig-4}     
\end{figure}

Figure~\ref{fig-4} (left) shows 2D angular correlations from high-multiplicity 200 GeV \pp\ collisions (first) and an inferred nonjet quadrupole amplitude vs multiplicity soft component $\bar \rho_s$ (second)~\cite{ppquad}. The quadrupole amplitude ($\propto$ number of correlated pairs) increases $\propto \bar \rho_s^3$ accurately over a thousand-fold amplitude increase. Referring to participant low-$x$ gluons with $N_{part} \propto \bar \rho_s$ and $N_{bin} \propto \bar \rho_s^2$ (dijet production) the \pp\ quadrupole then varies as $\propto  N_{part} N_{bin}$.

%%%%%%%%%%
%%%%%%%%%%

Figure~\ref{fig-4} (right) shows  2D angular correlations from mid-central 200 GeV \auau\ collisions, dominated by the NJ quadrupole, and the centrality trend for the quadrupole amplitude (divided by $\epsilon_{opt}^2$) vs {\em nucleon} participant pairs $N_{part}/2$~\cite{anomalous,v2ptb}. The solid line is consistent with the amplitude trend $\propto  N_{part} N_{bin} \epsilon_{opt}^2$. In both cases quadrupole amplitudes are inferred by model fits to 2D angular correlations that exclude any significant jet contribution. The simple \pp\ and \auau\ trends appear to be closely related, and there is no evidence for response to high densities and QGP formation. The \pp\ $\bar \rho_s^3$ trend especially suggests that the quadrupole arises from a three-gluon interaction and is not a manifestation of hydro expansion (flows).

\begin{figure}[h]
	\includegraphics[width=1.23in,height=1.2in]{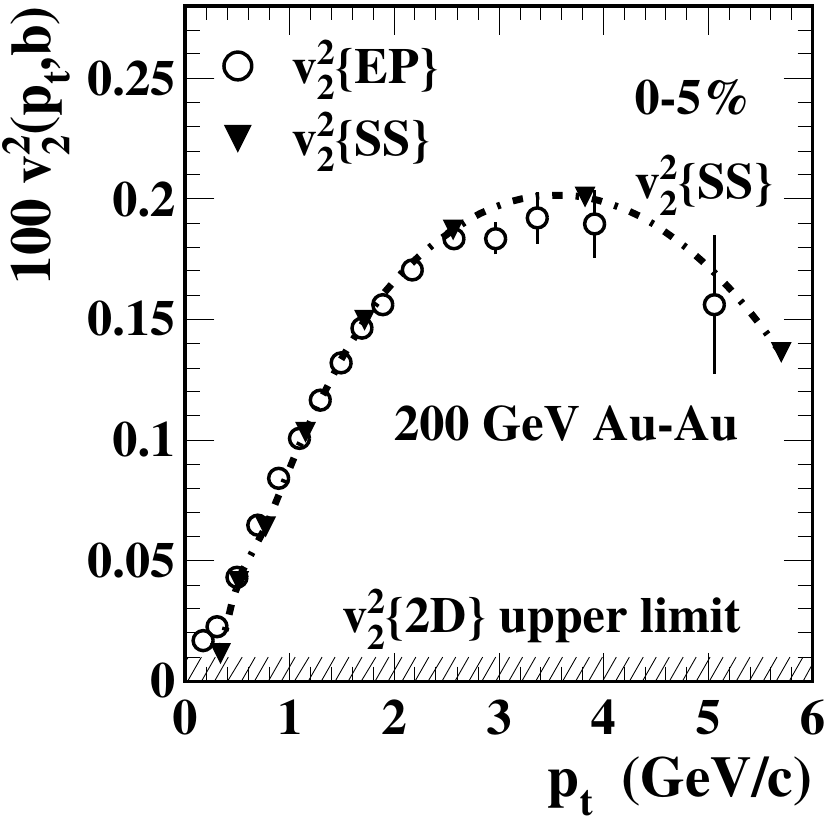}
	\includegraphics[width=1.25in,height=1.23in]{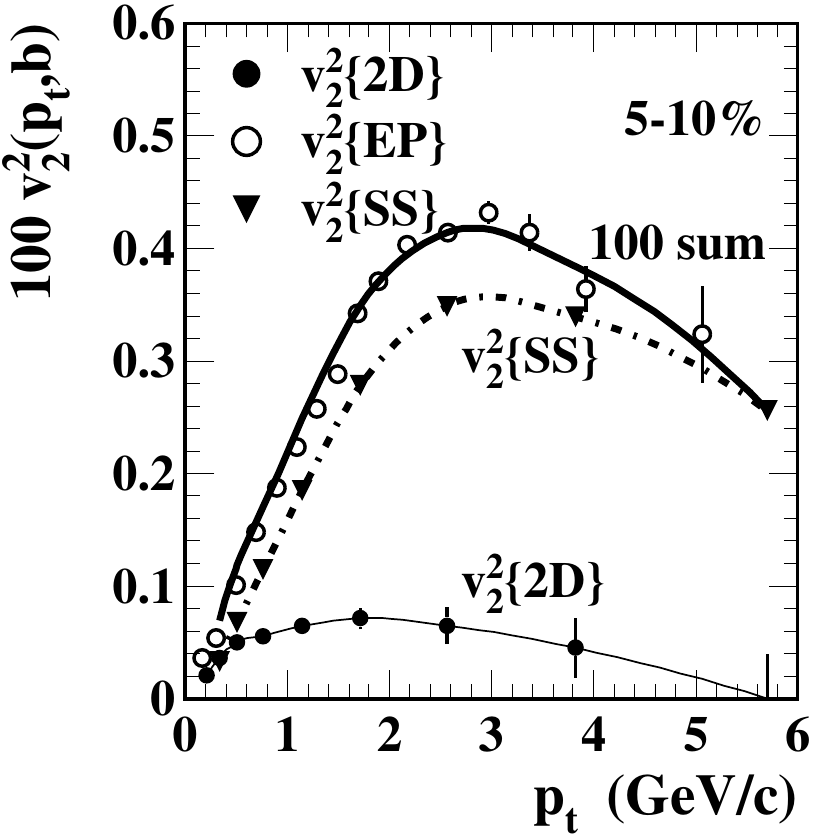}
	\includegraphics[width=1.25in,height=1.2in]{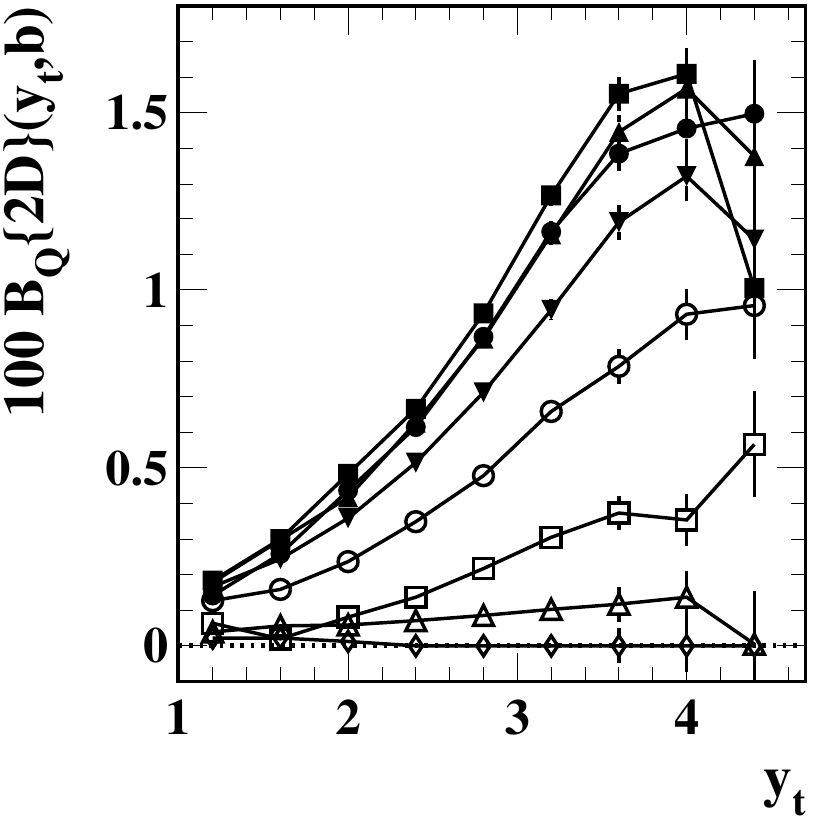}
	\includegraphics[width=1.25in,height=1.2in]{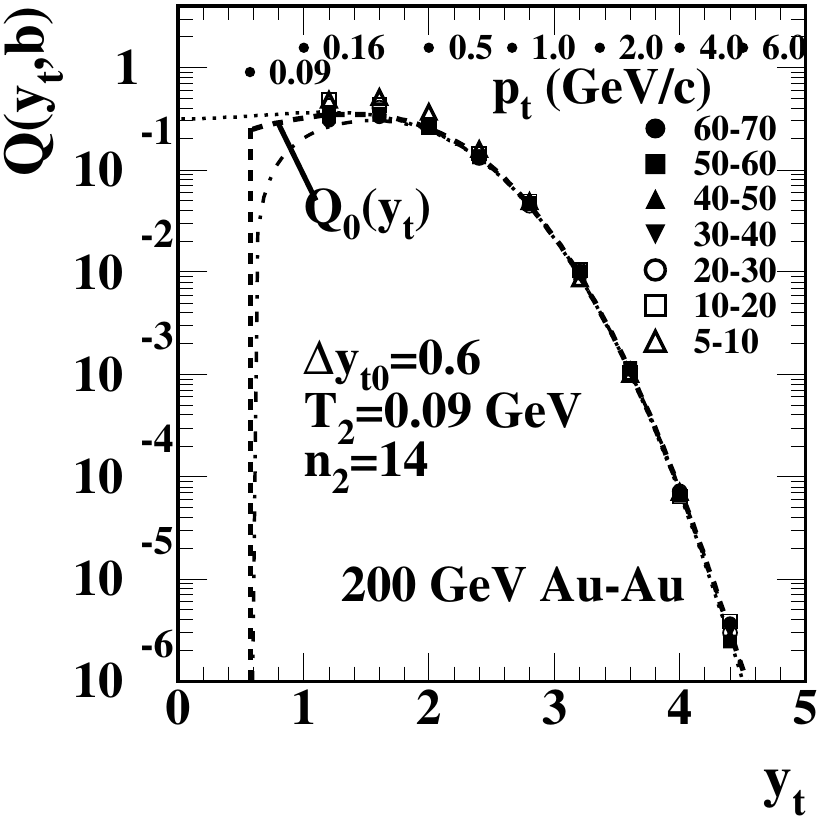}
	\caption{
		Left: $v_2^2\{\text{EP}\}(p_t)$ trend for 200 GeV \auau\ collisions.
		Right:  $v_2^2\{\text{2D}\}(p_t)$ trend for 200 GeV \auau\ collisions.
	}
	\label{fig-5}     
\end{figure}

Figure~\ref{fig-5} (left) shows $v_2\{\text{EP}\}$ (event-plane method) data (open circles) from more-central 200 GeV \auau\ collisions compared with $v_2\{\text{SS}\}$ derived from the same-side 2D jet peak (solid triangles)~\cite{v2ptb}. $v_2\{\text{2D}\}$ NJ quadrupole data (solid points) are also inferred from 2D model fits.
In more-central Au-Au collisions $v_2\{\text{EP}\}$ data are dominated by or entirely determined by jets -- i.e. as a Fourier component of the SS 2D jet peak projected onto 1D azimuth.

%%%%%%%%%%%%%%%%

%%%%%%%%%%
%%%%%%%%%%

Figure~\ref{fig-5} (right) shows quadrupole amplitude $B_Q\{\text{2D}\}$ derived from $v_2\{\text{2D}\}$ data for 200 GeV \auau\ collisions vs transverse rapidity \yt\ (first) and quadrupole spectra in the lab frame inferred from those data (see Fig.~\ref{fig-1}, left)~\cite{v2ptb}. The quadrupole spectrum shapes and the source boost $\Delta y_{t0} \approx 0.6$ derived from  $v_2\{\text{2D}\}$ data are independent of Au-Au centrality even for peripheral collisions. There is no evidence for the presence of a dense QCD medium.

%%%%%%%%%%%
\section{Do hydro models have any relation to real A-B collisions?}
\label{sec-6}

The apparent success of hydro models in describing spectra and angular-correlation data from an array of A-B collisions has been viewed as convincing evidence for QGP formation. Hydro descriptions of $x$-A data as reported in Ref.~\cite{nature} are invoked to buttress claims of QGP-droplet formation in those collision systems. However, such interpretations can be questioned.

\begin{figure}[h]
	\includegraphics[width=2.5in,height=1.18in]{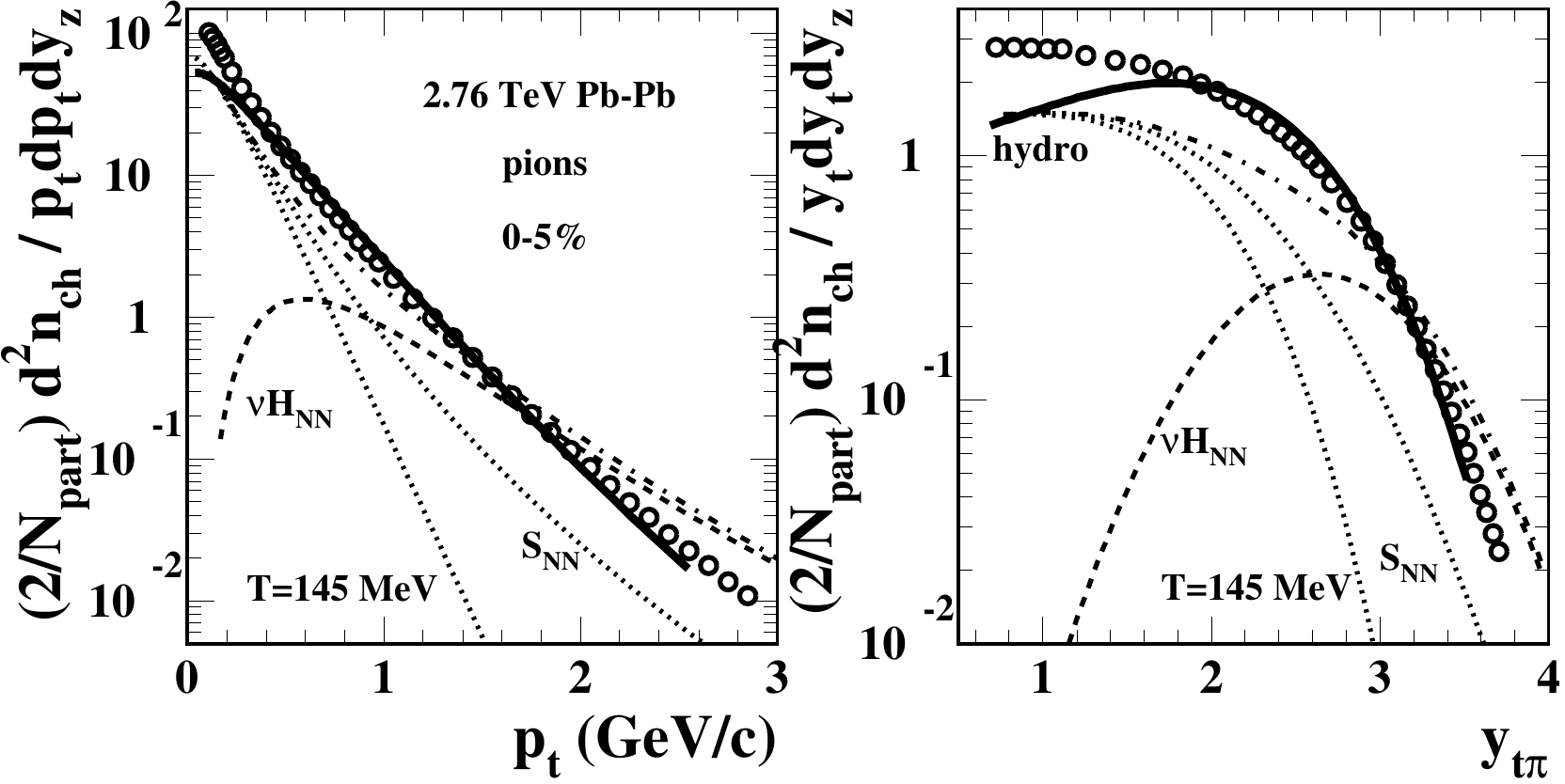}
	\includegraphics[width=1.25in,height=1.2in]{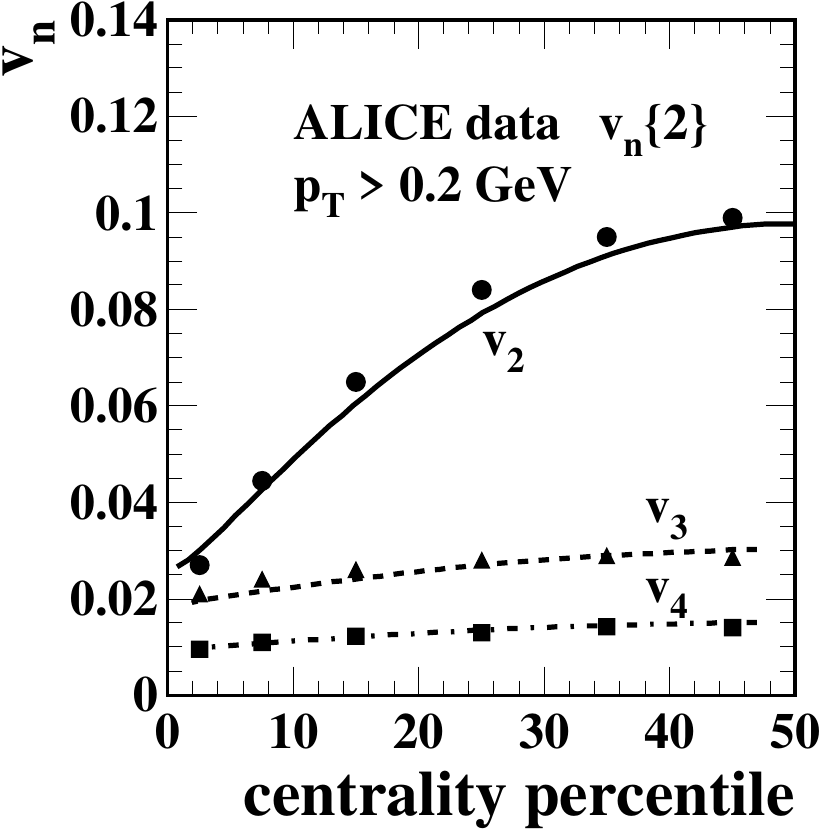}
	\includegraphics[width=1.25in,height=1.2in]{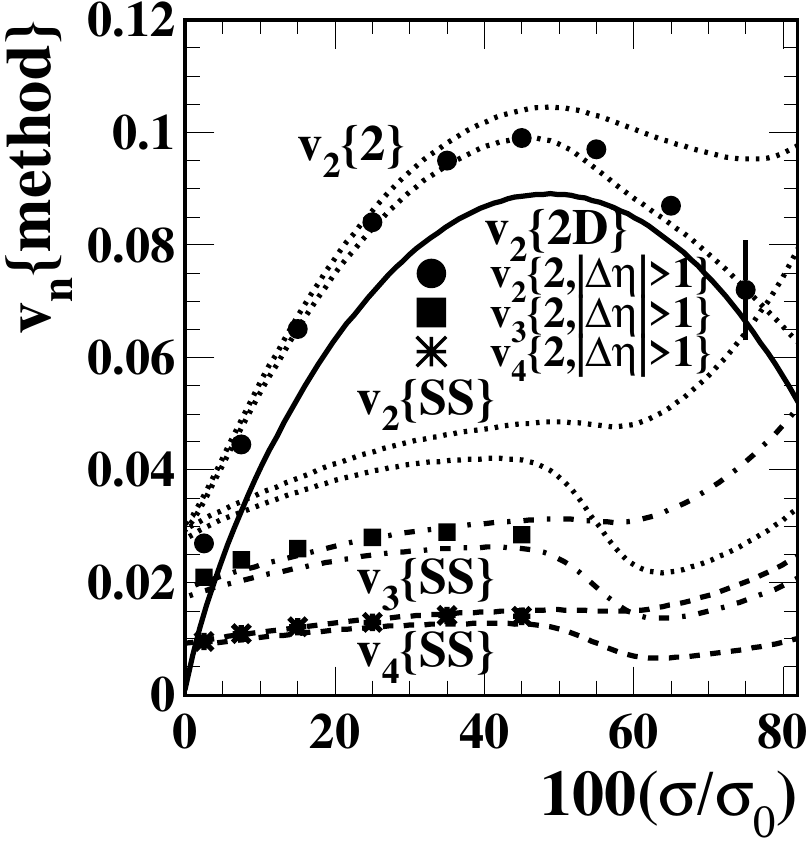}
	\caption{
		Left: Hydro theory compared to pion spectrum data from 2.76 TeV \pbpb\ collisions~\cite{galehydro2,tomnature}.
		Right:  Hydro theory compared to $v_2(b)$ centrality data from 2.76 TeV \pbpb\ collisions~\cite{galehydro2,multipoles}.
	}
	\label{fig-6}     
\end{figure}

Figure~\ref{fig-6} (left) shows hydro theory (solid curves)~\cite{galehydro2} compared to a pion spectrum from 0-5\% central 2.76 TeV \pbpb\ collisions (open circles). In the first panel, on linear \pt, the apparent agreement between theory and data at higher \pt\ is emphasized. However, in the second panel, on logarithmic \yt, the large discrepancy below \yt\ = 2 ($p_t = 0.5$ GeV/c) falsifies the hydro model. Strong suppression at lower \pt\ is expected from a hydro model (arising from source boost) but is not observed in data. Hydro theory in effect accommodates the large jet contribution $H_{NN}$ (TCM hard component) peaked near \yt\ = 2.7 ($p_t \approx 1$ GeV/c)~\cite{tomnature}.

%%%%%%%%%%
%%%%%%%%%%

Figure~\ref{fig-6} (right) shows hydro theory (curves in third panel)~\cite{galehydro2} compared to $v_n$ data vs centrality from  2.76 TeV \pbpb\ collisions (points). The hydro curves appear to describe the data well. However, the fourth panel shows an alternative description of the same ALICE data (albeit with additional points that are missing from the hydro treatment)~\cite{multipoles}. The dotted, dashed and dash-dotted curves in the fourth panel are {\em predictions} for $v_n$ derived from jet features inferred from model fits to 2D angular correlations~\cite{anomalous}. The solid curve represents the nonjet quadrupole feature obtained from the same fits. The step up in jet-related trends from peripheral to central collisions (near $\sigma / \sigma_0 \approx $ 0.5) corresponds to the ``sharp transition'' in jet structure occurring in Fig.~\ref{fig-1} (third panel) that relates to ``jet quenching.'' Especially for $v_3$ and $v_4$ the hydro model accommodates only jet-related correlation structure. One can conclude from these examples that at least some current hydro models are sufficiently complex and variable to accommodate data that may have no relation to flows. They are not predictive.

%%%%%%%%%%%
\section{Summary}

Sections 2-6 respond to five critical assumptions (expressed here in the form of questions) that form the basis for argument in a recent Nature letter introduced to claim formation of QGP droplets in small asymmetric $x$-A collisions. The evidence invoked for QGP formation in small collision systems presents a problem with two possible resolutions: (a) QGP formation is apparently a universal phenomenon in high-energy nuclear collisions, requiring novel theoretical approaches in response, or (b) certain ``signatures'' conventionally associated with QGP formation in \aa\ collisions are misinterpreted in any collision system and hydro descriptions of certain data features do not necessarily correspond to a flow mechanism, may actually be accommodating minimum-bias jet contributions. Experimental evidence presented in this talk responding to the five assumptions calls into question identification of certain data features with QGP formation and challenges hydro theory descriptions of data that have been interpreted to confirm the presents of flows.

The two-component (soft + hard) model (TCM) of hadron production provides a simple alternative description of hadron production in high-energy nuclear collisions that is consistent with basic QCD and with a broad array of jet measurements. The TCM offers methods and results that enable accurate distinctions among jet contributions (hard) and nonjet contributions (soft) to yields, spectra and two-particle correlations. In \pp\ and \pa\ collisions the TCM provides an exhaustive description of data with no need for exceptional mechanisms (e.g.\ no flows or jet quenching). In \aa\ collisions the TCM provides a stable and predictive {\em reference} against which deviations from linear superposition (e.g.\ jet modification) can be assessed quantitatively. A third, nonjet quadrupole, component with simple trends in \pp\ and \aa\ collisions appears to arise from an elementary QCD process (few-gluon interactions).

\end{document}